# Getting to know Classical Novae with Swift

Julian P. Osborne
University of Leicester
University road, Leicester, LE1 7RH, UK
opj@leicester.ac.uk

**Abstract**

Novae have been reported as transients for more than two thousand years. Their bright optical outbursts are the result of explosive nuclear burning of gas accreted from a binary companion onto a white dwarf. Novae containing a white dwarf close to the Chandrasekhar mass limit and accreting at a high rate are potentially the unknown progenitors of the type Ia supernovae used to measure the acceleration of the Universe. Swift X-ray observations have radically transformed our view of novae by providing dense monitoring throughout the outburst, revealing new phenomena in the super-soft X-rays from the still-burning white dwarf such as early extreme variability and half- to one-minute timescale quasi-periodic oscillations. The distinct evolution of this emission from the harder X-ray emission due to ejecta shocks has been clearly delineated. Soft X-ray observations allow the mass of the white dwarf, the mass burned and the mass ejected to be estimated. In combination with observations at other wavelengths, including the high spectral resolution observations of the large X-ray observatories, high resolution optical and radio imaging, radio monitoring, optical spectroscopy, and the detection of GeV gamma-ray emission from recent novae, models of the explosion have been tested and developed. I review nine novae for which Swift has made a significant impact; these have shown the signature of the components in the interacting binary system in addition to the white dwarf: the re-formed accretion disk, the companion star and its stellar wind.

## 1. Introduction

### 1.1. Classical novae

Classical and recurrent novae are due to the runaway nuclear burning of hydrogen-rich material slowly accreted on to a white dwarf from its binary companion. The recurrent novae are those classical novae that have been observed to undergo outburst more than once. Novae are thus cataclysmic variables, and they are supposed to represent a recurring phase in the long-term evolution of CVs towards a minimum orbital period[1]. Novae have been reported for more than two thousand years because of their very large optical outburst amplitude, which may be 7 - 15 optical magnitudes. At peak they can exceed the Eddington luminosity, and ejection of the outer layer of the WD is a natural consequence; indeed broad optical emission lines are commonly seen, implying velocities up to many thousand km.s$^{-1}$ in the very fastest fading novae. As a result of this ejection beautiful shells of glowing gas surround some novae. Useful overviews and introductions to novae have been published by Bode & Evans (1989, 2008), Warner (1995), Hellier (2001) and most recently by Shara (2014); Schaefer (2010) provides a full overview of recurrent novae.

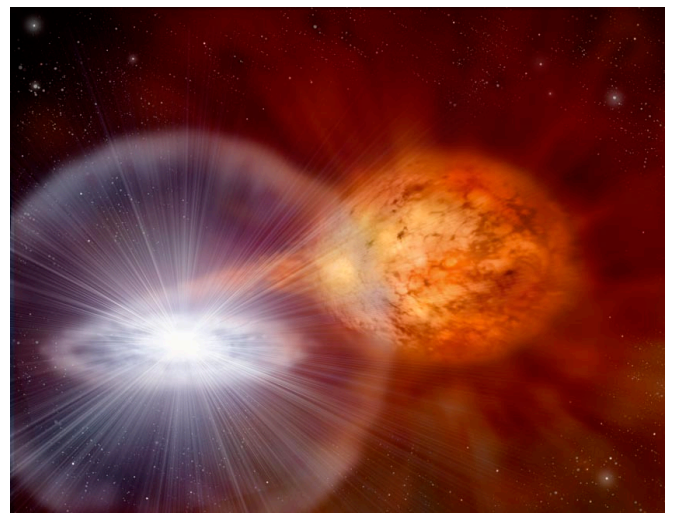

**Fig 1.** An artist's impression of the recurrent nova RS Oph (credit: David A. Hardy/PPARC).

Novae are sources of dust, and more significantly they are the primary producers of carbon-13, nitrogen-15 and oxygen-17 in the Galaxy (Starrfield et al. 2008). Novae also show a fascinating variety of behaviours in the optical band, with multiple light curve and spectral classifications made; some have multi-peaked outbursts, some have equatorial and polar ejecta. However, the greatest question facing us today is: *Are some novae progenitors of some type Ia supernovae?* The origin of these supernovae, which are used to measure the accelerating expansion of the Universe, is currently unknown. No candidate progenitor system appears to meet all the requirements (Wang & Han 2012; Maoz et al. 2014), thus leaving a major hole in this important cosmological result. Novae, as accreting white dwarfs, some of which are close to the Chandrasekhar mass limit, are natural candidates, although not all nova types are likely progenitors. Recent results suggest that the SNIa population is not as homogeneous as once thought (e.g. Milne et al. 2015), perhaps anyway allowing for multiple progenitor types.

---

[1] From now on I shall mean both classical and recurrent novae when I say novae. Of course these systems are distinct from the so-called 'X-ray novae', also known as 'soft X-ray transients', which are accretion events in a binary usually containing a black hole.

### 1.2. High-energy observations before Swift

X-ray detections of novae started with the EXOSAT observatory. These were able to show an intensity rise and fall after nova outbursts (Ögelman et al. 1987), but their limited spectral band-pass and resolution hindered unique physical interpretation. These early observations are discussed by Ögelman (1990).

ROSAT was able to provide much-improved soft X-ray observations, culminating in the 18-point light curve of the 1992 nova V1974 Cyg (Krautter et al. 1996). Rising and falling over 650 days, this initially hard source became the brightest super-soft source[2] ever seen. These two spectral components were respectively ascribed to shocks within the ejecta and residual nuclear burning on the white dwarf.

XMM-Newton and Chandra allowed continuous long-duration observations and the use of high spectral resolution gratings for the first time, although early observations suffered from unfortunate timing. Nova V1494 Aql in 2000 was the first for which a high quality soft spectrum was measured, although surprisingly Rohrbach et al. 2009 found this was dominated by emission lines. This object also showed unexpected flares and oscillations (Drake et al. 2003). Nova V4743 Sgr (2002) did exhibit the anticipated super-soft spectrum from the white dwarf 180 days after outburst, as well as a poorly-understood 22 min oscillation and flux decline from full intensity to zero in ~ 1 h (Ness et al. 2003). Comparison with model atmosphere spectra by Rauch et al. (2010) implied a WD mass of 1.1-1.2 $M_\odot$. The super-soft source in nova V5116 Sgr (2005) 20 months after outburst showed one and bit faint and bright states, with a recurrence consistent with the orbital period of 2.97 h, suggesting an eclipse by a re-formed accretion disk (Sala et al. 2008). Krautter (2008), Orio (2012) and Ness (2012) provide reviews covering this era.

### 1.3. The Swift advantage

The Swift satellite, launched into low earth orbit in November 2004, together with its associated communication and ground systems, is designed for optimal observations of gamma-ray bursts (Gehrels et al. 2004). With a payload of a large, wide-field coded mask 15-150 keV Burst Alert Telescope (BAT, Barthelmy et al. 2005), and sensitive co-aligned narrow field 0.3-10 keV X-ray and 170-600 nm UV/optical telescopes (XRT: Burrows et al. 2005. UVOT: Roming et al. 2005), Swift is very well suited to a wide range of astrophysical investigations. The key to unlocking the potential of this set of instruments is the highly autonomous, rapidly slewing nature of the spacecraft in combination with high sky availability, flexible scheduling of observations executed even within an hour or so from request, delivery of high-level data products a few hours after the observation, and an open and generous target of opportunity programme. This combination provides a radically new, and near-ideal facility for following up energetic fast transients discovered with other telescopes.

Swift has transformed observations of novae. Not only has it been able to map out in exquisite detail the full high-energy light curves of many novae from within a day of outburst, the superb low-latency data delivery and scheduling has allowed new phenomena to be explored in detail as they have occurred on hour to day timescales. In addition, Swift has enhanced the efficiency of other facilities, notably XMM-Newton and Chandra, by enabling them to target the important outburst evolution phases with confidence by providing contemporary intensity information; many such observations would not have been approved without the timely information from Swift.

### 2. Swift observations of novae

Swift has observed some 66 novae (to the end of April 2015), of which 39 have been detected in X-rays. Observations have started within a day of outburst discovery in some cases, and have extended to more than 14 yr after outburst for V723 Cas. A dozen novae have been sufficiently interesting to have had more than 100 ks spent on them. Swift observes Galactic novae, those in the Magellanic Clouds and in M31. Ness et al. (2007) published an early overview; the super-soft X-ray sample was described by Schwarz et al. (2011). There is not space here to describe all the Swift nova results; I will concentrate on some of the more intensively studied objects.

---

[2] Super-soft source (SSS) is an observational classification originally given to X-ray sources with essentially no detection above 0.5 keV, equivalent to a blackbody temperature of $kT \sim 30$ eV ($T \sim 3 \times 10^5$ K) and $L_X \sim 10^{38}$ erg.s$^{-1}$ (Hasinger 1994), although now including higher temperatures also.

## 2.1. RS Oph

The 2006 outburst of the recurrent nova RS Oph was the focus of unprecedented attention, with more than 40 papers in the refereed literature, and a dedicated conference (Evans et al. 2008). This symbiotic system (i.e. having a red giant secondary) has a 454-day orbital period and is embedded in the dense, slow-moving wind from the red giant. Well known from 7 previous eruptions, which occur every 20 yr or so, the outbursts of RS Oph are akin to small and rapidly evolving supernovae on account of the strong shock that the ejecta form with the dense wind. This was immediately evident from the highly absorbed hard X-ray spectrum seen by both Swift (Bode et al. 2006) and RXTE (Sokoloski et al. 2006) to cool and emerge from the overlying wind within a week or so of the outburst. Over this interval the shock was detected in the lowest energy channel of the Swift/BAT (14-25 keV), the only nova to have been clearly detected in this way so far. This shock was found to be bipolar with an equatorial ring in radio interferometry data by O'Brien et al. (2006), a model which was dramatically confirmed by direct imaging with HST 155 days after the outburst (Bode et al. 2007). This resulted in an expansion velocity measurement of ~5,600 km.s$^{-1}$ and the inference that the expanding nebula is shaped by the geometry of the wind in the binary system. Walder et al. (2008) support this view with a 3-D model of the circumstellar mass distribution before outburst.

The X-ray shock, modelled hydrodynamically by Vaytet et al. (2007) as a radiating system fed by an on-going fast wind from the white dwarf, rather than the previous single-impulse assumption, was found to have the observed velocity plateau-decline behaviour, requiring a low ejected mass (~$10^{-6} M_\odot$). Analysis of the Chandra grating X-ray shock emission line spectrum allows similar ejected mass values for enhanced metal abundances in the ejecta, as indicated by the absorption of these lines due to the red giant wind and explosion ejecta at day 14 (Drake et al. 2009). Further support comes from the comparison of synthetic spectra from the hydrodynamic shock model to those obtained by the Swift/XRT; good fits require an ejected mass of (2-5)x$10^{-7}$ $M_\odot$ and a nova fast wind velocity ~10,000 km.s$^{-1}$ (Vaytet et al. 2011). As an escape velocity, this points to a very high mass white dwarf, ~1.3 $M_\odot$, and thus a potential SNIa progenitor, however, the accretion rate over the outburst cycle is not well known, and the case is not yet proven. In addition, comparison of the velocities inferred from the X-ray shock temperature and IR emission line widths was shown to imply escape of TeV protons which dominated the shock cooling at early times (Tatischeff & Hernanz 2007).

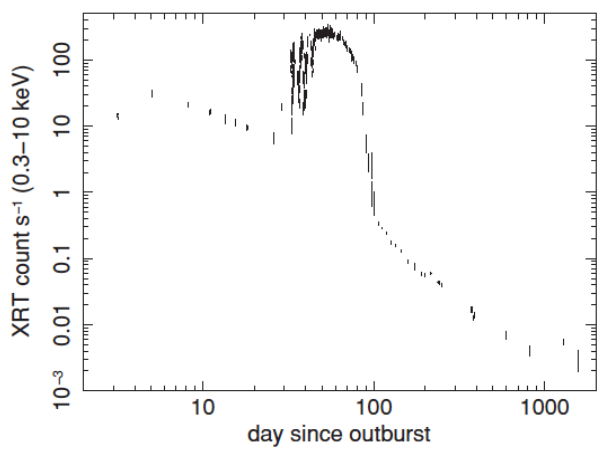

**Fig. 2.** Swift XRT 0.3-10 keV light curve of the 2006 outburst of RS Oph, the bright super-soft phase is clearly visible, from around days 30 to 100, on top of the slower hard X-ray evolution (from Osborne et al. 2011).

Twenty-six days after the outburst, the X-ray behaviour of RS Oph changed dramatically (see Fig. 2). The flux below 1 keV rose steeply in an extremely noisy fashion to day 46, with the Swift XRT count rate increasing by a factor of 30, after this it declined increasingly quickly so that the ratio of soft to hard flux was constant after day 120. During this time RS Oph became the brightest super-soft source ever seen. The temperature of the Eddington luminosity nova photosphere was seen to rise to 90 eV with the photosphere shrinking from the time of peak flux (Osborne et al. 2011). This is broadly what is expected from nova models, in which the reducing nova wind density allows the photosphere to shrink down towards the white dwarf radius, while the decreasing optical depth exposes increasingly hotter layers. Nuclear burning of residual fuel probably ended on day 80, when the white dwarf temperature was seen to decline rapidly. This nice picture is to be treated with some caution however, as the soft X-ray spectrum was evidently complex, and the photosphere appeared to shrink at roughly constant temperature for much of the super-soft phase (~45-75 days after outburst). The early extreme variability of the super-soft emission, a factor >10 in a few hours, is not seen in the contemporary Swift UVOT measurements, and so is likely due to absorption by ejecta blobs passing through the line of sight.

A significant surprise during the super-soft phase of RS Oph was the detection of a roughly 35 second quasi-periodic oscillation (Osborne et al. 2011), see Fig. 3. With an amplitude up to 10%, this modulation was noticeably more coherent in the second half of the super-soft phase than the first. No similar periodicity had been seen from any nova prior to this discovery. The period can be due to the spin of the white dwarf (as other accreting white dwarfs are known to have a similar spin period), although the causes of the modulation and of its low coherence remain to be explained. Alternatively, the modulation may be related to a nuclear burning pressure oscillation, although the early models that do exist suggest long growth times for this instability. This oscillation was also seen strongly by XMM-Newton (Ness et al. 2015).

initially hard and eventually became so again, as was the case for RS Oph. However from around day 30 to 150 the X-ray count rate below 1.5 keV dominated that at higher energies; while the hard X-rays showed a slow rise and a slower fall, the soft X-ray count rate rose by a factor of ~300 in less than 10 days to peak on day 40 (Page et al. 2010, see Fig. 4). During the super-soft phase blackbody temperatures of $kT \sim 90$ eV (and model atmosphere temperatures of >80 eV) were measured, pointing to a very high white dwarf mass (e.g. Wolf et al. 2013). This nova showed neither the extreme variability at onset, nor the quasi-periodic oscillation of the soft X-rays seen in RS Oph, but a rise in photospheric temperature like that of RS Oph was seen as the super-soft source became apparent. Page et al. and Takei et al. (2011) suggest, on the basis of X-ray flickering observations, that accretion onto the white dwarf resumed <57 and around 50 days after outburst respectively; setting a constraint on the viscous disk spreading timescale.

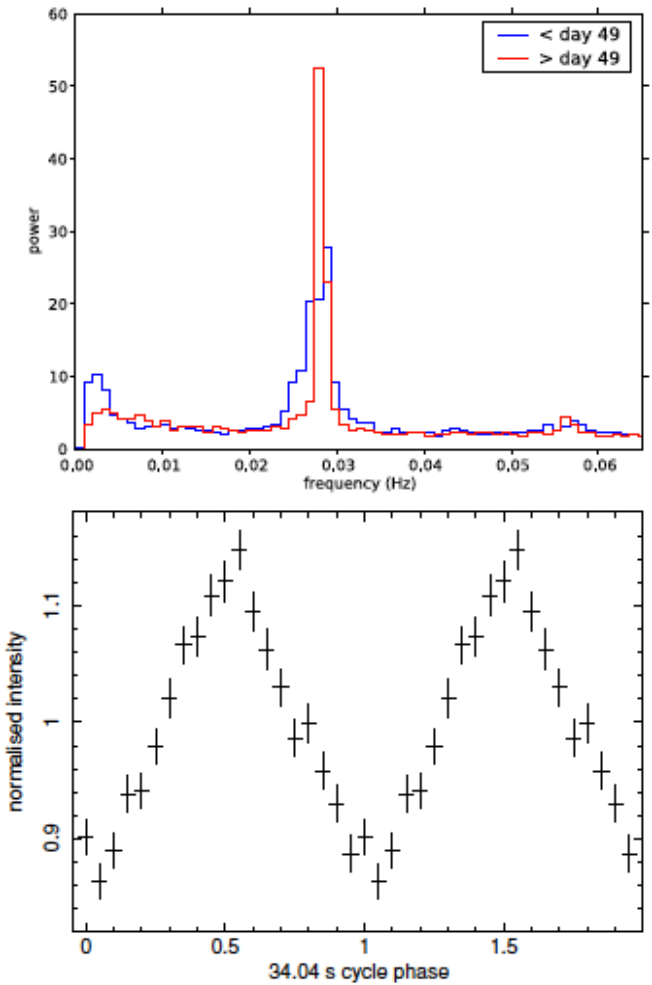

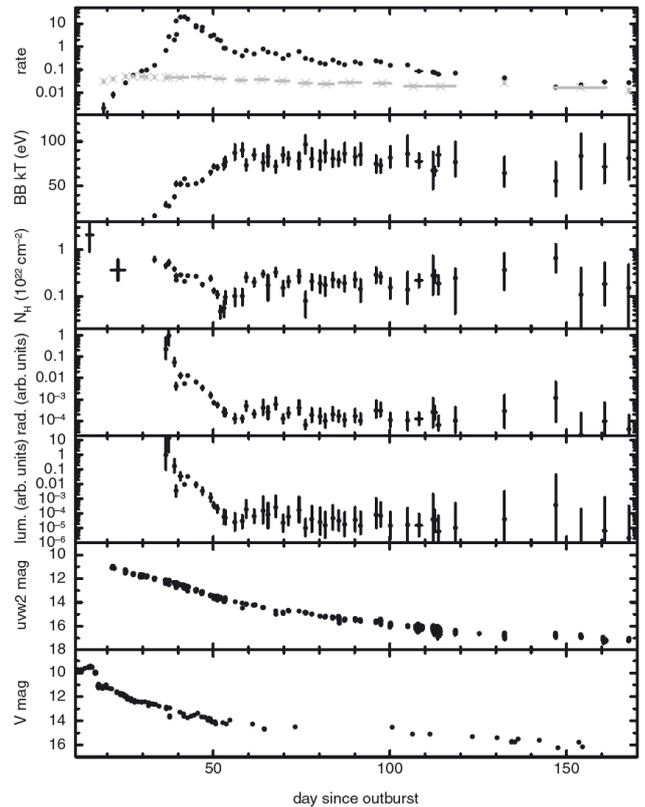

**Fig. 4.** Swift XRT, UVOT and AAVSO V-band light curve of V2491 Cyg (0.3-1.5 keV X-ray count rate in black, 1.5-10 keV in grey). Also shown are time-resolved blackbody spectral fit parameters showing a temperature rise and radius decline during the super-soft phase (from Page et al. 2010).

**Fig. 3.** Top - Fourier power spectra of the super-soft X-ray emission of RS Oph showing greater coherence of the quasi-period oscillation at later times. Bottom - The shape of the periodic modulation on day 37.2. (From Osborne et al. 2011).

### 2.2. V2491 Cyg

Swift observed the 2008 nova V2491 Cyg within a day of outburst, an unprecedentedly fast response time. Observed over 250 days, the X-rays were

This object is only the second to have been seen in X-rays prior to outburst, having a luminosity in the range $1-40 \times 10^{34}$ erg.s$^{-1}$ up to 3 months beforehand (Ibarra et al. 2009). Together with the high ejecta velocity of this nova (~5,000 km.s$^{-1}$) and very rapid optical decline, suggesting a high mass white dwarf, the high inter-outburst luminosity suggests that V2491 Cyg is a recurrent nova with a likely outburst cycle of a few $10^2$ to a few $10^4$ yr (Page et al. 2010). The high white dwarf mass proposal is supported by the remarkable similarity of the X-ray grating spectra of V2491 Cyg and RS Oph (Ness et al. 2011), and the XMM-Newton grating analysis pointing to an ONe white dwarf (Pinto et al. 2012). 2MASS data and optical colours demonstrate that the secondary star in this system is not a red giant, but a sub-giant (Darnley et al. 2011), and thus in contrast to RS Oph the orbital period is just a few days, and there is no strong secondary star wind. The origin of the V2491 Cyg hard X-rays seen early in the outburst must therefore be different, and presumably they arise in internal shocks within the ejecta. Darnley's revised distance to this nova of 14 kpc implies a still higher quiescent luminosity, and thus a recurrence timescale <100 yr (although no recurrence has been found). In a review of novae that may be recurrent Pagnotta & Schaefer (2014) consider the case for a recurrent classification of V2491 Cyg to be 'conflicted' because of its high outburst amplitude and optical light curve showing a very unusual secondary maximum.

### 2.3. V407 Cyg

This object heralded one of the great surprises in the modern era of nova studies: it was detected by the Fermi/LAT at energies >100 MeV (The Fermi-LAT Collaboration 2010). The gamma-ray flux was undetected before outburst, peaked 3-4 days after outburst and faded below detection by day 16. During this interval the X-ray flux was low, only rising significantly immediately afterwards. V407 Cyg has a Mira-type red giant secondary, and hence a dense slow-moving wind against which the rapid nova ejecta may shock. The LAT emission is consistent with Fermi acceleration of particles in the shock; the emitted spectrum being due to either proton-proton or electron inverse Compton processes, with the latter favoured in later consideration by Martin & Dubus (2013). The unusual nature of the secondary star in this system suggested that high-energy gamma-ray emission from novae would be rare.

The ejecta-wind shock was studied in detail by Shore et al. (2011), who showed that the X-ray emission, although never bright, required a complex, cooling model, and declining absorption over 60 days as the shock expanded through the wind. Super-soft emission from the white dwarf was not strongly evident, although Nelson et al. (2012) obtain better spectral fits when this is included. RS Oph is the natural comparison object even though the orbital period of V407 Cyg is decades rather than just over a year; optical emission lines indicated an asymmetric shock development, similar to RS Oph. Hydrodynamic modelling by Orlando & Drake (2012) shows that the X-ray emission comes predominantly from the dense region of converging shocks behind the secondary star, see Fig. 5. Their best fit to the X-ray light curve requires a circumbinary density enhancement, due to gravitational focussing of the red giant wind by the white dwarf, an explosion energy of $\sim 2\times 10^{44}$ erg and an ejecta mass $\sim 2\times 10^{-7}$ $M_\odot$. The dense Mira wind was constrained to $\sim 2\times 10^{-6}$ $M_\odot.yr^{-1}$ by VLA radio observations (Chomiuk et al. 2012); the rise and fall of the lower frequency flux over many months was found to be due to the increasing ionisation of the wind followed by heating from the nova shock, the first radio transient recognised to be formed in this way.

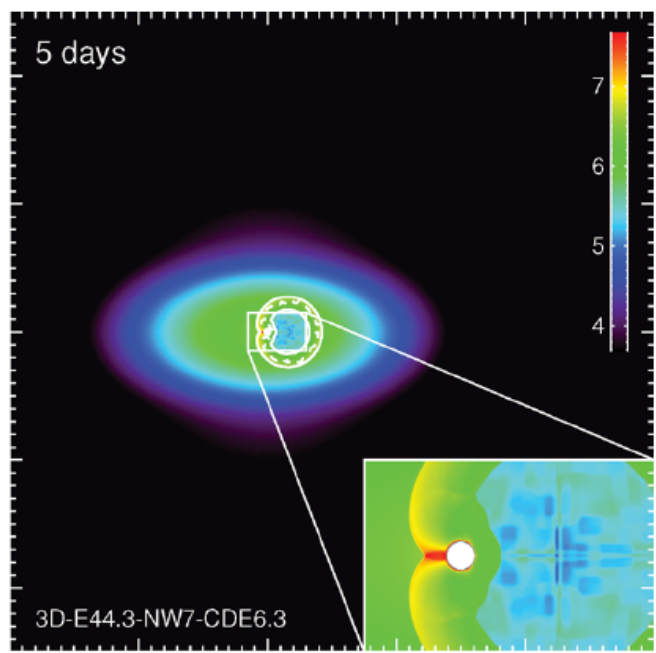

**Fig. 5.** Hydrodynamic model of the V407 Cyg ejecta 5 days after outburst assuming a disk-like circumbinary density enhancement. The colour scale shows log density (cm$^{-3}$), the white dashed contour encloses ejecta, the white solid contour plasma at T>1 MK. The large ticks are 100 AU. The inset shows the effect of the shock passage around the secondary star (from Orlando & Drake 2012).

### 2.4. HV Cet

Discovered by the Catalina Sky Survey an unknown time after outburst, HV Cet [3] had an optical spectrum of a late nova with a strong emission line from the high excitation energy [Ne V] (Preito et al. 2008), suggesting the presence of X-rays and a very high mass oxygen-neon white dwarf. Swift observations revealed that the X-ray emission lasted to >160 days after discovery, to be entirely below 1 keV, and also to be strongly modulated at a 1.77 day period (Beardmore et al. 2010). Some Swift observations took place at high cadence, once every 1.5 h orbit, revealing how the modulation varied with time, and the in-phase modulation of the 192.8 nm UV flux (also seen in AAVSO optical data; Fig. 6). Fitting the XRT spectra with an NLTE model atmosphere resulted in temperatures between 60 - 80 eV (which did not vary on the 1.77 day period), and luminosities around a factor 300 less than the expected Eddington value (Beardmore et al. 2012).

---

[3] = CSS081007:030559+054715

Taking the 1.77 day period to be orbital, the picture that emerges is one of a high inclination system in which the white dwarf is obscured by a raised accretion disk rim, the height of which is largest where the stream from the secondary impacts. The observed X-rays are then the result of optically thin electron scattering into the line of sight by gas close to the white dwarf. The UV luminosity is close to Eddington, and is supposed to come from the inner face of the raised accretion disk rim. This object is unusual both for being at high Galactic latitude, and for the composition of the white dwarf which implies a mass above 1.16 $M_\odot$.

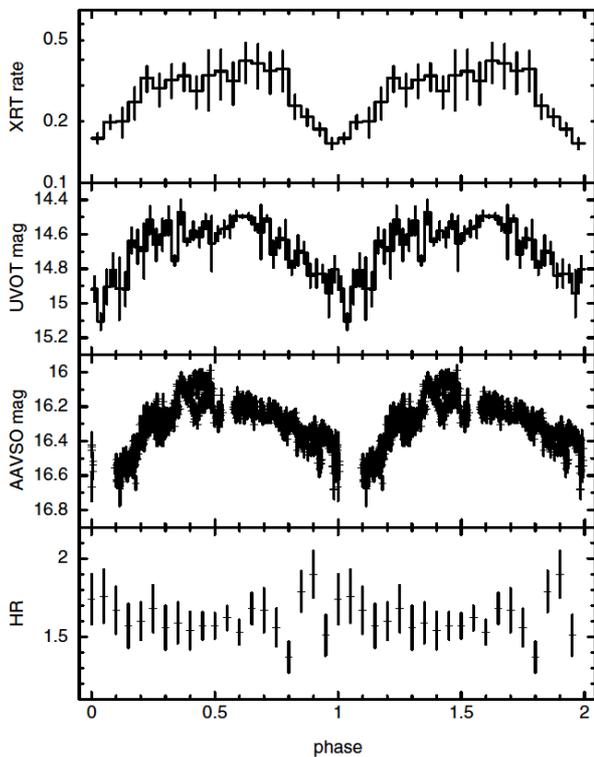

**Fig. 6.** HV Cet data from Swift XRT (count rate and hardness), UVOT, and AAVSO, 71-84 days after discovery, folded on the 1.77 day period (from Beardmore et al. 2012).

### 2.5. T Pyx

T Pyx was the first recognised recurrent nova, but it has many unusual and apparently paradoxical properties. It is the only nova below the CV period gap, having $P_{orb}$ = 1.83 h, yet it has had a recurrence interval of ~20 yr. The low accretion rate expected at such a short period does not provide sufficient fuel for such frequent outbursts. X-ray heating of the secondary may increase the accretion rate, but such X-rays have not been seen (in spite of the low orbital inclination of the system). The sixth outburst of T Pyx in 2011 occurred after a 44-yr break, the longer interval likely due to a secular decrease in accretion rate. Schaefer et al. (2013) give a comprehensive overview.

Because T Pyx is both so well known and so odd, the 2011 eruption was spectacularly well covered at all wavelengths. Chomiuk et al (2014a) brought together the Swift data, with VLA radio data and the results of optical spectroscopy. A long plateau to around day 100 dominates the optical light curve, whereas the X-ray and radio flux do not substantially pick up until around the same time (see Fig. 7). The X-ray spectrum consists of relatively oxygen-rich optically thin emission and a cool blackbody component that fades rapidly from day 145-200 whilst increasing in temperature from ~30 to ~60 eV. The optical plateau is ascribed to a stalled initial low mass ejection. Photospheric shrinkage and ejecta clearing allow the super-soft X-ray emission to become visible around day 100. At about the same time a much more massive ejection at high velocity on day ~75 causes the two ejecta shells to collide, so giving rise to the hard X-ray and radio fluxes which peak at around day ~150. The ejection velocities observed, in combination with the time of peak radio flux rule out an ejecta-circumstellar shock. The super-soft X-ray turn-on time suggests a high ejecta mass, of a few x $10^{-5}$ $M_\odot$, much higher than expected for a recurrent nova; the white dwarf temperature requires a modest white dwarf mass of ~ 1.0 $M_\odot$ (a conclusion also reached by Tofflemire et al. (2013) on the basis of X-ray grating observations). In spite of this relatively self-consistent picture of the T Pyx eruption, the origin of the second very energetic mass loss from the white dwarf remains unclear.

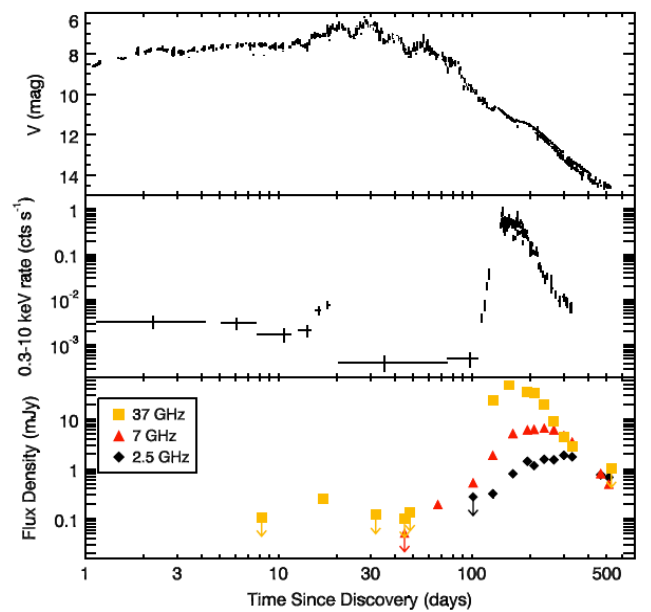

**Fig. 7.** Optical, X-ray and radio evolution of T Pyx (from Chomiuk et al. 2014a).

## 2.6. V959 Mon

Nova V959 Mon in 2012 was the third nova to be seen by the Fermi-LAT at gamma-ray energies above 100 MeV (Ackermann et al. 2014), and was the first to be discovered using this energy band. As the second gamma-ray nova was little studied, this eruption provided a good opportunity to examine the model proposed for the first gamma-ray nova; their gamma-ray properties appear quite similar.

Page et al. (2013) show that this object has a coherent flux modulation in the X-ray, UV and I bands with a period of 7.1 h (see also Munari et al. 2013). This modulation is the cause of most of the rapid variability in the soft X-ray flux, see Fig. 8. Such a modulation most naturally arises from an accretion disk rim expanded at the point of impact of the stream from the companion; a high binary inclination is required, and this is demonstrated through bipolar ejecta modelling by Ribiero et al. (2013). This short orbital period rules out a red giant secondary star, as is present in the first gamma-ray nova V407 Cyg, and thus also rules out the existence of a dense circumstellar wind. Such a wind was a key component of the initial model for gamma-ray emission, although Martin & Dubus (2013) later showed that the gamma-ray spectrum and flux evolution of V407 Cyg were better fit with a density enhancement around the white dwarf than with a red giant wind; this density enhancement may be a natural consequence of an accretion disk.

Model atmosphere fits to the super-soft X-ray spectrum require a white dwarf mass ~1.25 $M_\odot$, consistent with the oxygen-neon classification ascribed by Shore et al. (2013).

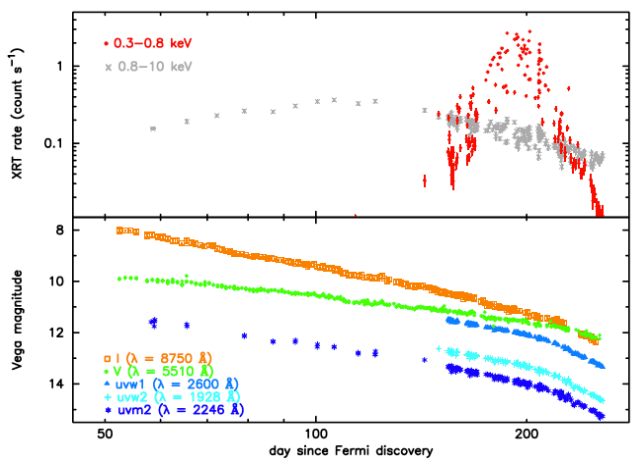

**Fig. 8.** V959 Mon: Soft and hard XRT, and IR-UV light curves (from Page et al. 2013). The high-frequency variability in the soft X-ray flux (red) is partly due to the 7.1-h period.

## 2.7. V745 Sco

The 2014 eruption of this recurrent nova was the first to be observed with the sensitive hard X-ray NuSTAR telescopes. Swift also monitored it from just 3.7 h of its first announcement for almost a year (Page et al. 2015). V745 Sco is a very fast nova, in that it declines from peak by two magnitudes in just 2 days, the fastest known in our Galaxy. It contains a late M type giant, and so a dense wind surrounds the system. It was detected by Fermi-LAT as a weak gamma-ray source.

The simultaneous NuSTAR-Swift dataset taken 10 days into the outburst constrains the shock emission exceptionally well (Fig. 9). Orio et al. (2015) find that the spectrum above 1 keV is dominated by locally absorbed shock emission in collisional ionisation equilibrium at kT = 2.8 keV with iron at half-solar abundance; the white dwarf photosphere dominates at lower energies. They find no evidence for a power law spectral component, as would be expected from Comptonisation of gamma-ray emission, limiting the gamma-ray flux at this time to around an order of magnitude less than the level detected in the first 2 days.

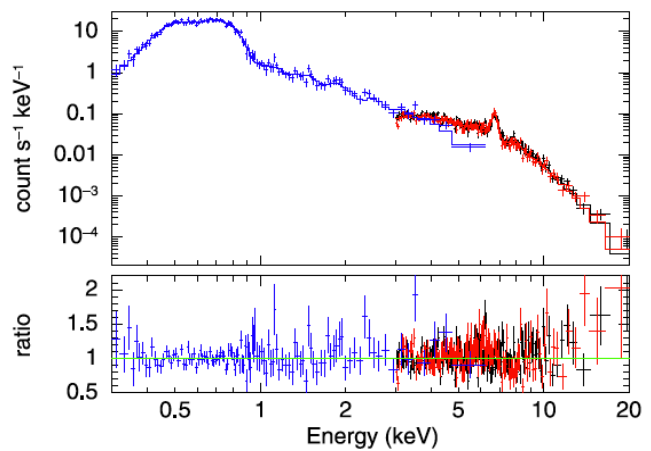

**Fig. 9.** V745 Sco observed X-ray spectra from Swift and NuSTAR fit with a sub-solar iron abundance shock and a white dwarf photosphere model (from Orio et al. 2015).

## 2.8. LMC 2012-03a

Nova LMC 2012-03a had a similarly fast optical decline to V745 Sco, and a very fast ejecta expansion velocity, ~5,000 km.s$^{-1}$. Few Magellanic Cloud novae have been well studied so far, but nova LMC 2012 was extensively observed by Swift, Chandra, HST-STIS and with optical spectroscopy from the ground. Super-soft X-ray emission was seen between days ~13 and 50 with the high model atmosphere temperature of kT = 86 eV (Schwarz et al. 2015); all of these properties point towards a very high mass white dwarf, probably > 1.3 $M_\odot$. Only a very low hard X-ray flux was observed, implying a shock luminosity three orders of magnitude below that typically seen from novae. Swift UVOT data revealed a period of 19.2 h, which was not seen in X-rays.

With the good data coverage, Schwarz et al. (2015) were able to create Cloudy models of the photoionisation caused by the very hot white dwarf at 7.5, 29.5, 42 and 57 days after discovery (the nova V-band maximum occurred at 0.3 days). Initially, no X-rays were detected because the photosphere is 10% of the binary separation (from the period observed) and so the temperature is well below the X-ray regime. The optical emission is due to $1.6 \times 10^{-6}$ $M_\odot$ of ejecta at this time, however this does not explain the later optical emission, which is a hundred times brighter than the ejecta would be then. The UV modulation points to illumination of the secondary star, and the Cloudy models show that this fits the observed emission well (see Fig. 10). The accretion rate of the near-Chandrasekhar mass white dwarf in a 19.2 h binary is $10^{-8}$ $M_\odot.yr^{-1}$, so that the accreted envelope would reach ignition pressure in just ~60 yr. For the observed super-soft duration the mass burned is a small fraction of the mass ejected, which itself is a small fraction of the mass accreted. The high mass white dwarf in this system is growing in mass in this current phase.

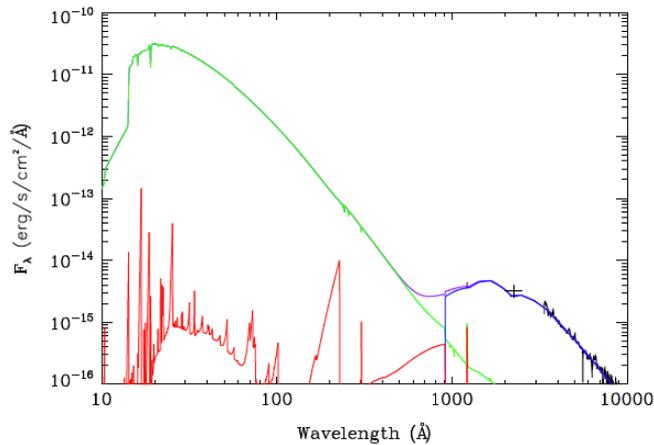

**Fig.10.** Nova LMC 2012 best fitting Cloudy model for day 42. The white dwarf is shown in green, the secondary star blue, the ejecta red. The UV-optical emission (black) is dominated by the illuminated secondary rather than the ejecta (from Schwarz et al. 2015).

### 2.9. M31 N2008-12a

Our nearest comparable neighbour galaxy, M31, is a great place to look for novae. M31 offers a common distance of 780 kpc, modest foreground absorption, a high surface density in a compact area, and the prospect of exploring various different galactic locations. M31 has been the subject of a number of surveys capable of detecting novae efficiently; the MPE M31 nova catalogue [4] contains almost a thousand objects. Early XMM-Newton observations found a nova-type super-soft X-ray source (Osborne et al. 2001) that King et al. (2002) found to be accretion-fed. Continued X-ray monitoring with XMM-Newton has continued to yield a substantial harvest of new novae (e.g. Henze et al. 2014a).

The M31 nova 2008-12a is one of the most remarkable and surprising known: in 2013 the fifth outburst was discovered, which indicated a nova recurrence interval of just one year (Darnley et al. 2014; Tang et al. 2014). The shortest Galactic nova recurrence timescale, that of U Sco, is 8 yr. The ultra-short eruption pattern of 2008-12a requires a very high mass white dwarf, as also suggested by the very fast optical decline rate of 2 magnitudes in just 4 days, and a high accretion rate. The very rapid super-soft X-ray turn-on and the very high blackbody temperature seen by Henze et al. (2014b) and Tang et al. (2014) are consistent with this view. The annual eruptions of this nova were confirmed by the observation of the predicted nova event in 2014 by Henze et al. (2015), for which Swift revealed the highly consistent behaviour of the super-soft X-ray source between outbursts (see Fig. 11). The 6-day super-soft X-ray turn-on suggests a very low ejected mass of $2-3 \times 10^{-8}$ $M_\odot$; the overall evolution of this is well fit by a model for a 1.377 $M_\odot$ white dwarf. Kato et al. (2014) show that the shortest possible nova recurrence period is about two months for a non-rotating 1.38 $M_\odot$ white dwarf.

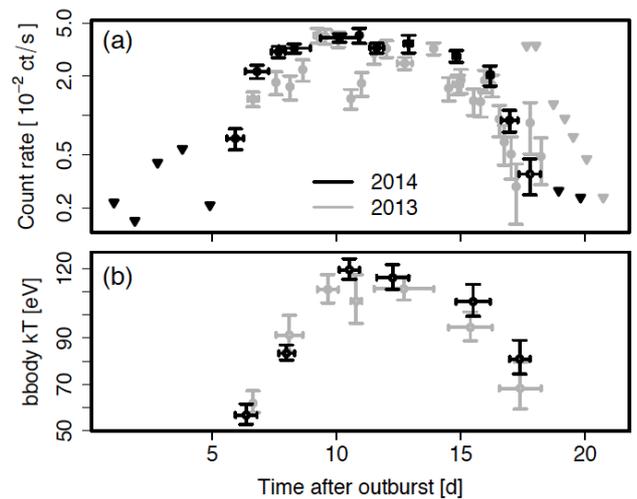

**Fig. 11.** Super-soft X-ray emission seen by Swift from consecutive outbursts in 2013 and 2014 of the remarkable recurrent nova M31N 2008-12a. Triangles show upper limits. (From Henze et al. , in press).

---

[4] http://www.mpe.mpg.de/~m31novae/opt/m31/

## 3. Overview and open questions

In this review I have given illustrations of how Swift has driven forwards our appreciation of classical and recurrent novae. Prior to Swift, X-ray observations had either poor spectral or temporal coverage. Swift has brought full temporal coverage of the outbursts, sometimes starting within hours of discovery. Swift can be scheduled at short notice according to the behaviour of the source, allowing observations every 1.5 hours. Swift allows spectroscopy of both the hard X-ray shocks and of the evolving photospheric emission and its faster variations. The high availability and responsiveness of Swift allows other facilities to have observations scheduled with confidence in the current state of the object, thus allowing a greater number to be made. The resulting coordinated multi-facility examination of novae has naturally enabled a much closer probing of the physical processes at work in these iconic objects.

After the early Swift overview by Ness et al. (2007), Schwarz et al. (2011) reviewed the properties of 52 Galactic and Magellanic Cloud novae. This large collection of results showed important correlations: fast optical decline with high ejecta velocity, and early turn-on and turn-off of the super-soft X-ray source with higher ejecta velocity (see Fig. 12); while a previously suspected X-ray turn-off time correlation with orbital period was shown not to hold. UV flux may be correlated, anti-correlated, or uncorrelated with the X-ray flux. The durations of the super-soft phase in these novae had a median value of 1.4 years, significantly longer than that seen in M31; this is most likely to be a selection effect, due to the way in which the Swift observations are pursued.

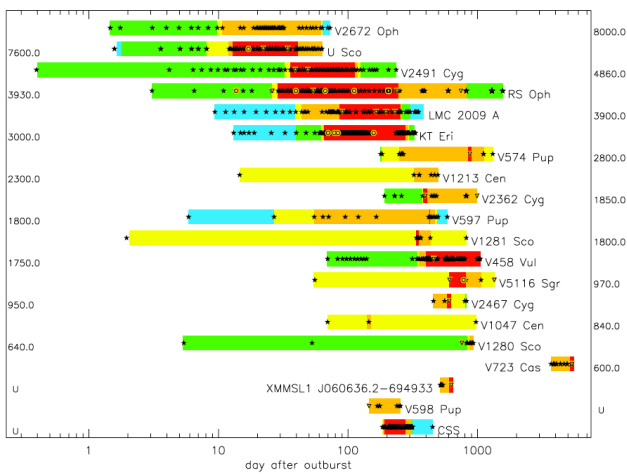

**Fig. 12.** The pattern of X-ray emission in novae observed by Swift, ordered top to bottom by high to low optical emission line FWHM. Observations are shown by stars, intervals are colour coded by X-ray spectral state: blue = undetected; green = hard; yellow = intermediate; orange = most likely soft; red = soft (from Schwarz et al. 2011).

Henze et al. (2011, 2014a) found similar correlations in the M31 novae; they also showed that the temperature of the white dwarf is much higher for the novae with earlier super-soft X-ray turn-off times.

These correlations can be understood in terms of the basic nova model (e.g. Starrfield et al. 1974, Iben & Tutukov 1989, Sala & Hernanz 2005, Yaron et al. 2005, Hachisu & Kato 2006, Wolf et al. 2013), in which the impulsive nuclear burning of accreted material on the white dwarf results in the prompt ejection and continued expansion of its outer layer, followed by Eddington-limited burning of residual hydrogen-rich fuel until this is exhausted. As the ejecta expand away from the white dwarf its density declines, so that the photosphere shrinks back towards the normal white dwarf radius; because this occurs at constant luminosity the photospheric temperature rises while this is occurring. Higher mass white dwarfs have a smaller radius, and thus a higher escape velocity. They also have deeper gravitational potential wells; burning occurs at higher temperatures, and the accreted gas reaches ignition pressure at a lower accreted mass, giving rise to lower ejecta mass from the more massive white dwarfs.

The combination of observations and the models allows values to be inferred for the white dwarf mass, and for the mass of the material accreted, burned and ejected. These are important considerations when considering novae as potential SNIa progenitors, as white dwarfs near the Chandrasekhar mass limit may be growing in mass. Other observational constraints apply to the question of novae as SNIa progenitors, not least the search for evidence of the secondary star in the early light curve and of evidence of long-term accretion prior to the SN (e.g. Olling et al. 2015; Cao, et al. 2015; Broersen et al. 2014; Graur et al. 2014). In addition, mass-growing white dwarfs of the recurrent novae may suffer dominating mass loss in much less frequent classical nova eruptions (e.g. T Pyx; Schaefer et al. 2013). The question remains unresolved so far.

Naturally, the derivation of physical parameter values depends on a correct interpretation of the mechanisms at work. The rise of a visible super-soft X-ray source is ascribed to a combination of the shrinking photosphere of a residual-fuel burning white dwarf and the thinning ejecta screen, but it may also be that renewed accretion is playing a role here. Ness et al. (2008) suggested that accretion was feeding the very long-lived super-soft X-ray emission of nova V723 Cas, and it is evident that the persistent close binary super-soft sources (Kahabka & van den Heuvel 2006), such as Cal 83 and Cal 87, are continuously powered by nuclear burning of accreting gas. In addition, we are increasingly seeing, with Swift and other facilities, evidence of the reformed accretion disk soon after the nova eruption (e.g. Beardmore et al. 2012, Page et al. 2013, Worters et al. 2007, Ness et al. 2012, Ness et al. 2013). Related to this suggestion of accretion disk fed super-soft emission in novae, Alexander et al. (2011) propose that the outbursts of RS Oph are due to disk mass transfer instability induced accretion events, which cause immediate nuclear burning.

There remain many areas ripe for further work. Some of these are:

1) Similar to the quasi-periodic oscillations seen in the super-soft X-rays from RS Oph, Ness et al. (2013) detail QPOs seen by XMM-Newton in the novae KT Eri (35 s), V339 Del (54 s) and LMC 2009a (33 s); also the persistent super-soft source Cal 83 has a QPO at 67 s. Beardmore et al. (2015) present the much longer Swift observation results. Potentially related to white dwarf spin or burning-induced oscillations, a modern investigation of instabilities in and above the nuclear generation zone is required to accept or reject this hypothesis, and perhaps to exploit the information the modulation may provide.

2) Swift has provided X-ray spectra, although at CCD resolution, throughout the evolution of the super-soft white dwarf photosphere. XMM-Newton and Chandra have provided high-resolution X-ray grating spectra in continuous multi-hour observations. Full specification model atmospheres for novae are not easy to create, as they have to cover the extended atmosphere, the velocity field, and the range of potential abundances. Currently available model atmospheres are either incomplete, or are highly tuned to specific circumstances so that they are not generally applicable. van Rossum (2012) made significant progress, but the full parameter space remains to be explored, and the great potential of the observations is yet to be realised.

3) The shock physics and the gamma-ray emission of novae is a highly active area now. Chomiuk et al. (2014b) proposed that the collision between the initial slow-moving ejecta in the binary equatorial plane and the later, much faster, white dwarf wind is the region in which the particle acceleration occurs to give rise to the radio synchrotron and gamma-rays observed. That work, derived from their high-resolution radio observations of V959 Mon as well as X-ray results, was supported by VLA observations of nova V1723 Aql (Weston et al. 2015). Metzger et al. (2015) take this further, arguing that a significant fraction of optical emission is powered by absorbed X-ray radiating shocks (thus not by the previously assumed thermal emission from the white dwarf). They suggest that hard X-ray observations by NuSTAR of gamma-ray novae at the time of optical peak may detect the shock emission [5], while the maximum particle energy can be constrained by TeV gamma-ray observations.

4) The final area of work I highlight is simply that of the continued detection and multi-wavelength observation of new novae in our Galaxy, in the Magellanic Clouds and in M31. Each of these galaxies offers specific advantages in the study of novae, and in each we are far from exhausting, or understanding, the variety of observational characteristics that they show. Novae share significant physics with supernovae, and advances in understanding each can help with the other, and eventually to knowledge of the progenitors of the Ia supernovae that are used to measure the acceleration of the Universe.

**Acknowledgements**

The success of Swift is due to the small, dedicated instrument, mission operations and data analysis software teams in the USA, the UK and Italy, following the highly effective leadership of the Principal Investigator, Neil Gehrels; it has been my pleasure to be a part of this, and to acknowledge this great team. I also acknowledge the skill and achievement of the University of Leicester XRT camera hardware team, led pre-launch by Alan Wells, from which the current Leicester team has evolved. The work described here has been mostly undertaken by the Swift nova-CV group, an open and loose collaboration that I have enjoyed co-ordinating with the significant help of Kim Page: http://www.swift.ac.uk/nova-cv The Swift project at the University of Leicester is funded by the UK Space Agency.

---

[5] Unfortunately the NuSTAR observation of V745 Sco occurred too late to make a useful test.